\documentstyle[twocolumn,prl,aps]{revtex}
\begin{document}
\title{Hybridization Mechanism for Cohesion \\
of Cd-based Quasicrystals}
\author{Yasushi Ishii}
\address{Department of Physics, Chuo University, Kasuga, Tokyo 112-8551, Japan}
\author{Takeo Fujiwara}
\address{Department of Applied Physics, University of Tokyo,  Hongo,
Tokyo 113-8654, Japan} 
\maketitle

\begin{abstract}
Cohesion mechanism of cubic approximant crystals of newly discovered binary
quasicrystals, Cd$_6$M (M=Yb and Ca), are studied theoretically. 
It is found that stabilization due to alloying is obtained 
if M is an element with low-lying unoccupied $d$ states.
This leads to conclusion that the cohesion of the Cd-based compounds 
is due to the hybridization of the $d$ states of Yb and Ca
with a wide $sp$ band.
Although a diameter of the Fermi sphere coincides with the strong 
Bragg peaks for Cd-Yb and Cd-Ca, the Hume-Rothery mechanism
does not play a principal role in the stability
because neither distinct pseudogap nor stabilization
due to alloying is obtained for isostructural Cd-Mg.   
In addition to the electronic origin, matching of the atomic size is 
very crucial for the quasicrystal formation of the Cd-based compounds.
It is suggested that the glue atoms, which do not participate in
the icosahedral cluster, play an important role in stabilization
of the compound.

\end{abstract}
Electronic structures of aluminum-based quasicrystals (QC)
have been studied so far
for many kinds of approximant crystals~\cite{review}.
The most characteristic feature in
the electronic structures of approximant crystals and probably QC is
a pseudogap in the density of states (DOS) at the Fermi
level.   It is believed that origin of the pseudogap is mainly due to 
the Brillouin-zone(BZ)-Fermi-sphere(FS) interaction
(the Hume-Rothery mechanism),
and presumably secondly due to the $s$-$d$ hybridization~\cite{allicu,friedel}.
In the Hume-Rothery mechanism, a strong interference
of electronic waves at ${\bf k}$ and ${\bf k+G}$ induces the
pseudogap near the Fermi level where ${\bf G}$
is the reciprocal lattice vector giving strong Bragg scattering and
satisfying $|{\bf G}| \approx 2k_F$.
This interference effect is more efficient in QC than in approximant
crystals because of sphericity of distribution of {\bf G}.
It is empirically known that stable QC is obtained if
the average number of valence electrons per atom, which is
usually denoted as $e/a$,  is close to either 1.7 or 2.1. 
The QC without transition elements,
such as Al-Li-Cu and Zn-Mg-RE (RE: rare earth),  belong to
a family of QC with $e/a=2.1$,  for which a diameter of the FS,
$2k_F$, calculated from the electron density is close to
the (222100) and (311111) reciprocal lattice vectors.

Very recently stable icosahedral QC have been found in binary 
Cd-Yb and Cd-Ca systems~\cite{tsai,guo,takakura}.
The quasicrystalline phases are identified
as {\it unknown} phases in the phase diagrams, 
Cd$_{5.7}$Yb and Cd$_{17}$Ca$_3$ \cite{phase}.
Cubic crystalline phases, Cd$_6$M (M=Yb, Ca),  are obtained in
the composition near the quasicrystalline one.    The isostructural
cubic alloys are also obtained for other systems with M=Sr~\cite{cdca},
Y~\cite{cdy} and most of rare-earth elements~\cite{cdre} 
although the icosahedral QC is realized only for M=Yb and Ca.

The Cd-based QC is not only the first stable binary QC but
also a unique system in many aspects.
First, the local atomic structure conjectured from that in the cubic 
phase is similar to the Al-Mn family of QC with $e/a=1.7$ 
although it contains no transition elements
as is discussed by Takakura {\it et al}~\cite{takakura}.
The core structure of an icosahedral cluster is also very unique 
as explained below.
Secondly, unlike the conventional QC without transition elements, 
$e/a$ for Cd-Yb and Cd-Ca alloys is exactly 2
and the icosahedral phase is not obtained for Cd-M 
with trivalent metals, M~\cite{guo}.
Finally, although $2k_F$ calculated from the electron density is close to 
the (222100) and (311111) reciprocal lattice vectors,
only Cd-Yb and Cd-Ca forms QC among the isostructural Cd-based alloys.
It is important to examine the mechanisms other than the BZ-FS
interaction, such as the $sp-d$ hybridization effects and 
the atomic diameter effects. 
In this article we shall investigate the electronic structures of
the cubic Cd$_6$Yb and Cd$_6$Ca crystals to study mechanism 
for cohesion of the newly discovered QC.   We also present the
electronic structures of isostructural compounds Cd$_6$Sr and Cd$_6$Mg
although Cd$_6$Mg is hypothetical.

Cd$_6$Yb is a body centered
cubic crystal with space group $Im\bar3$, which contains 168 atoms
(176 sites) in a cubic cell 
with lattice parameter $a=15.638$ \AA\ ~\cite{cdyb}.
Although detailed structural analysis for Cd$_6$Ca has not been made,
Cd$_6$Ca is believed to be isostructural to Cd$_6$Yb
with a lattice parameter $a=15.680$ \AA\ ~\cite{cdca}.
In Cd$_6$Yb, the cluster is placed at the corner and the body center
of a cubic cell~\cite{takakura}.   A core of the cluster is an atomic
shell of non-icosahedral symmetry as in a pseudo-Mackay cluster
in Al-Pd-Mn~\cite{alpdmn}.   Four Cd atoms are placed 
at vertices of a small cube with occupancy probability 0.5.
Because Cd atom is not small enough to occupy neighboring vertices,
the central shell of Cd atoms may be of a tetrahedral shape.  
The tetrahedral symmetry is a subgroup symmetry of the icosahedral one
and the symmetry axes of the tetrahedral core coincide with those of
outer icosahedral shells.  
Therefore, we presume that the tetrahedral core is very probable.
The second and third
atomic shells are a dodecahedron of 20 Cd atoms and an icosahedron of 
twelve Yb atoms, respectively.    The fourth shell is a Cd 
icosidodecahedron obtained by placing 30 Cd atoms on the edge of the Yb
icosahedron.

To avoid the fractional occupation of  cadmium atoms
in the electronic structure calculations, 
a tetrahedral cluster of four Cd atoms, instead of a cubic one,  
is placed at the corner and the body-center of the cubic unit cell.
We assume the same atomic positions as those obtained for Cd$_6$Yb
for all the compounds with an appropriate lattice constant
shown in Table \ref{table1}.
Calculation is done with the tight-binding linear muffin-tin orbitals 
(TB-LMTO) method in the atomic-sphere approximation (ASA)~\cite{lmto}.

In Fig.~\ref{dos1}(a), we show the total DOS for Cd$_6$Yb.
A narrow band at about $-0.8$ [Ryd] is the Cd-$4d$ band and does
not contribute to cohesion. 
Another narrow peak just below the Fermi level
is attributed to the Yb-$4f$ states.   
Since the narrow $4f$ band is almost filled,
Yb is divalent as is Ca.
This is consistent with measurements of magnetic 
susceptibility~\cite{cdyb}.

A shallow dip in the DOS is seen between the $4f$ band and an unoccupied 
peak at $0.0$-$0.1$ [Ryd], which is made from the Yb-$5d$ states.
By checking decomposition of the DOS to partial waves, 
we find that the occupied states below the dip are 
predominantly made from the Cd-$5p$ states
except for the narrow $f$ band.   This is a qualitatively different
feature from pure Cd metal, in which the $s$ and $p$ states contribute
equally to the states near the Fermi Level.  
We speculate that hybridization
of the Cd-$5p$ and Yb-$5d$ orbitals makes the bonding
orbitals below the Fermi level leading to the dip (or the pseudogap)
in the DOS.

The total DOS for Cd$_6$Ca, Cd$_6$Sr and Cd$_6$Mg are shown 
in Fig.~\ref{dos1}(b)-(d).
Band width, position of the $d$-band and other characteristics of Cd$_6$Yb
are also found in Cd$_6$Ca and Cd$_6$Sr except for the narrow $f$ band
in Cd$_6$Yb.   For Cd$_6$Mg where magnesium has no low-lying unoccupied
$d$ state near the Fermi level,  on the other hand,
the shallow dip near the Fermi level vanishes.
Therefore we can say that hybridization of the $d$ 
states near the Fermi level is essential for the dip formation 
in the Cd-based compounds.   
This should be contrasted with the cases of
Al-Li-Cu~\cite{allicu} and Zn-Mg-Y~\cite{znmgy}, where the $sp$-$d$ 
hybridization is rather minor in the pseudogap formation. 
This fact was confirmed by checking that the pseudogap
does not vanish even if Cu in Al-Li-Cu and Y in Zn-Mg-Y are replaced 
with the elements without the $d$ states, Al and Mg, respectively.

Calculated cohesive energies per atom at a fixed
lattice constant are shown in Table \ref{table1}.   
The cohesive energy for Cd$_6$Mg at the optimal lattice parameter 
is similar to that for pure Cd (1.48 [eV/atom] for fcc with $a$=4.45 [\AA]) and 
smaller by about 15 \% than those for the other Cd$_6$M compounds 
with the low-lying $d$ band near the Fermi level.
Therefore the hybridization of the $d$ state with a wide $sp$ band
certainly contributes to stabilizing the Cd$_6$M compounds.
Although there have been a lot of conjectures that the $sp$-$d$ hybridization
is important for stabilizing the QC~\cite{friedel,trambly},
the present calculation is the {\it direct} demonstration that 
the $sp$-$d$ hybridization induces the pseudogap or the dip in the DOS
and lowers the structural energy of approximant crystals of QC.

A diameter of the FS is calculated from the electron density as
$2k_F = 2.75$ [\AA$^{-1}$], which is very close to
the (222100) and (311111) Bragg scatterings at 2.79 and 2.90 [\AA$^{-1}$],
respectively, for the icosahedral Cd-Yb~\cite{guo}.
This seems to support the Hume-Rothery mechanism, in which
the BZ-FS interaction induces the pseudogap near the Fermi energy.   
However,  there is neither distinct dip (or pseudogap) in the DOS
nor additional stabilization due to alloying
for isostructural and isovalent Cd$_6$Mg, which
has a similar structure factor to the other Cd$_6$M compounds.
Thus we should say that the interference effect associated with
the strong peaks in the structure factor is not of primary importance 
for the stability of the Cd-based compounds.

For the stable QC without transition elements,
$e/a$ is usually close to 2.1.   This empirical rule
for $e/a$ is not satisfied for the present
Cd-based compounds because all the elements are divalent
and hence $e/a$ is exactly 2~\cite{guo}.
Although a quantitative argument is difficult in the ASA,
a ratio of the $d$-symmetric state to the $sp$-symmetric 
one in the Yb/Ca atomic sphere increases by alloying 
in comparison with pure Yb and Ca.   This reminds us of
the negative valence trend of transition elements 
in the Al-Mn family of QC~\cite{friedel,trambly}.
The electron transfer to the $d$ state together with
the strong hybridization effects implies that the
electronic structure of the Cd-based compounds 
is very different from that of the conventional QC
without transition elements.  

The Fermi levels of Cd$_6$Yb and  Cd$_6$Ca are pinned at the shoulder
of the occupied band, not at the minimum of the DOS.   
Nevertheless reasonable amount of energetic stabilization is 
obtained for the cubic Cd$_6$M compounds as shown above.
This is because the occupied states just below the dip are bonding 
orbitals, whose levels are lowered by alloy formation.
One expects, however, that the Fermi level is shifted to 
the minimum of the DOS, yielding the larger cohesive energy 
by substituting trivalent atoms for Yb and Ca.   
In fact, we have checked that the Fermi level for
the isostructural Cd$_6$Y is located at the minimum of the DOS and
the cohesive energy per atom is as large as 2.15 [eV/atom].

So far we have seen that the cubic Cd$_6$M
compounds are stabilized if M is an element with low-lying unoccupied $d$ states.
It is reasonable to believe that the same mechanism works also in QC
because the local atomic structure would be similar.
Nevertheless, no quasicrystalline phase is obtained for Cd-Sr
at the composition close to Cd$_6$M.
As Guo {\it et al.}~\cite{guo} have already pointed out,
the atomic radius of Yb and Ca 
may be more suitable for quasiperiodic arrangements of atoms.
The atomic radii calculated from the lattice constants
of pure systems are listed in Table \ref{table1}~\cite{comment}.
The atomic radius of Sr is larger by 10 \% than those of Yb and Ca.
It may be rather surprising that the Cd-Sr compound  has the isostructural 
cubic phase with Cd$_6$Yb and  Cd$_6$Ca in spite of such difference.

Guo {\it et al.}~\cite{guo} also argued that alloying of Cd with trivalent 
elements  may not favor the QC formation 
because $e/a$ shifts apart from 2.
The electronic energy could, however, gain 
by the substitution of trivalent elements as mentioned above.
The atomic radii of the trivalent rare-earth elements 
and Y are considerably small (1.75-1.87 [\AA]) in comparison with
the divalent elements, Yb and Ca.   Matching of the atomic size is 
certainly important for the QC formation in the binary systems.
We should point out here that only europium is exceptional in this context:
Although Eu is divalent and the atomic radius is similar to Ca and Yb,
the QC is not obtained for the Cd-Eu alloy.   This is an
open question to be answered in the future.


Finally, to observe the electronic structure in connection with
the shell structure of the icosahedral cluster, we calculate the local density
of states (LDOS) for atoms in the individual atomic shells in the cluster.
In Fig.~\ref{ldos}, we show the LDOS for the central Cd$_{4}$ tetrahedron
and that for the glue Cd atoms, which do not participate in 
the icosahedral cluster.   A midgap peak appears at around 
$-0.1$ [Ryd] for the LDOS at the central Cd$_{4}$ tetrahedron
and the hybridization effect is not remarkable at the center of the cluster.
On the other hand,  the dip above the Fermi level is clearly 
seen in the LDOS  at the glue-atom sites and the DOS below the 
dip is enhanced.  We speculate that the Cd-$3p$ states on the glue atoms are 
strongly hybridized with the Yb-$5d$ states to favor the cluster packing.

We have studied the electronic structures of the cubic Cd$_6$M
approximant crystals of the newly discovered binary QC. 
It is found that stabilization due to alloying is obtained 
if M is an element with low-lying unoccupied $d$ states.
This leads to conclusion that the cohesion of the Cd-based compounds 
is certainly due to the hybridization of the $d$ states of Yb and Ca
with a wide $sp$ band.   
Although a shallow dip in the DOS appears near the Fermi level
and a diameter of the FS coincides with the strong Bragg peaks
for Cd-Yb and Cd-Ca, the BZ-FS interaction does not play a principal role
in the Cd-based systems 
because neither distinct dip nor additional stabilization
due to alloying is obtained for isostructural Cd-Mg. 
In order to exclude definitely the Hume-Rothery mechanism,
further studies are obviously needed, however.
In spite of similar electronic structures for the cubic Cd$_6$M (except Mg),
only Cd-Yb and Cd-Ca form QC.   
It is conjectured that matching of the atomic size is 
crucial for long-range quasiperiodic packing of the clusters.

The authors would like to thank A.P.Tsai for valuable discussions. 
This work is partly supported by Core Research in Environmental 
Science and Technology, Japan Science and Technology Cooperation.


\begin{table}
\caption{Cohesive energies per atom $\Delta E$ 
and lattice constants $a$
for cubic Cd$_6$M together with atomic radii of M (M=Yb, Ca, Sr and Mg). 
The lattice parameter for hypothetical Cd$_6$Mg is a calculated
optimal one.
}
\label{table1}
\begin{center}
\begin{tabular}{@{\hspace{\tabcolsep}\extracolsep{\fill}}ccccc} 
M & Yb & Ca & Sr & Mg \\ 
\hline
$\Delta E$ [eV/atom] & 1.68 & 1.75 & 1.69 & 1.47 \\
$a$ [\AA]   &15.638 & 15.680 & 16.040  & 15.22 \\
atomic radius [\AA]   & 1.94 & 1.97 & 2.15 & 1.60 \\ 
\end{tabular}
\end{center}
\end{table}
\begin{figure}
\caption{Total (black)  and partial (green: $p$-wave,
red : $d$-wave) density of states for (a) Cd$_6$Yb, (b) Cd$_6$Ca, 
(c) Cd$_6$Sr and (d) Cd$_6$Mg.}
\label{dos1}
\end{figure}
\begin{figure}
\caption{Local density of states (LDOS) for Cd$_6$Yb 
at the central Cd$_{4}$ tetrahedron (a) and that at the glue Cd atoms (b),
The Cd-$3d$  components are eliminated from the LDOS.}
\label{ldos}
\end{figure}

\begin{references}
\bibitem{review} T.Fujiwara,  {\it Physical Properties of Quasicrystals},
edited by Z.M.Stadnik (Springer,1998), p.169.
%
\bibitem{allicu} T.Fujiwara and T.Yokokawa, Phys. Rev. Lett. {\bf 66},
333 (1991).
%
\bibitem{friedel}  J.Friedel, Helvetica Physica Acta, {\bf 61}, 538 (1988).
%
\bibitem{tsai} A.P.Tsai, J.Q.Guo, E.Abe, H.Takakura and T.J.Sato
Nature, {\bf 408}, 537 (2000). 
%
\bibitem{guo} J.Q.Guo, E.Abe and A.P.Tsai, 
Phys. Rev. B, {\bf 62}, 14605 (2000).
%
\bibitem{takakura} H.Takakura, J.Q.Guo and A.P.Tsai,
Philos. Mag. Lett. {\bf 81}, 411 (2001).
%
\bibitem{phase}  T.B.Massalski, H.Okamoto, P.R.Subramanian 
and K. Karcprzak, {\it Binary Alloy Phase Diagrams}, 
2nd edition (ASM International, 1990).
%
\bibitem{cdca} G.Bruzzone, Gazzeta Chimica Italiana, {\bf 102}, 234 (1972).
%
\bibitem{cdy} A.C.Larson and D.T.Cromer, 
Acta Crystallogr. , {\bf B27}, 1875 (1971).
%
\bibitem{cdre}  I.Johnson, R.Schablaska, B.Tani and K.Anderson,
Trans. Met. Soc. AIME, {\bf 230}, 1485 (1964).
%
\bibitem{cdyb} A.Palenzona,
J. Less-Common Metals, {\bf 25}, 367 (1971).
%
\bibitem{alpdmn}  M.Boudard, M. de Boissieu, C.Janot, G.Heger,
C.Beeli, H.-U.Nissen, H.Vincent, R.Ibberson, M.Audier and J.M.Dubois,
J. Phys. Condens. Matter {\bf 4}, 10149 (1992).
%
\bibitem{lmto}  O.K.Andersen, O.Jepsen and D.Gl\"otzel,
{\it Highlights in Condensed Matter Theory}, edited by
F.Bassani, F.Fumi and M.P.Tosi (North-Holland, New York,1985), p.59. 
%
\bibitem{znmgy} Y.Ishii and T.Fujiwara, in preparation.
%
\bibitem{trambly} G.Trambly de Laissardi\`ere, D.Nguyen Manth,
L. Magaud, J.P. Julien, F. Cyrot-Lackmann and D.Mayou,
Phys. Rev. B {\bf 52}, 7920 (1995).
%
\bibitem{comment} The atomic radii may be different in
alloys from those in pure systems if the electron transfer to the 
{\it localized} $d$  states due to alloying is significant.  
\end{references}
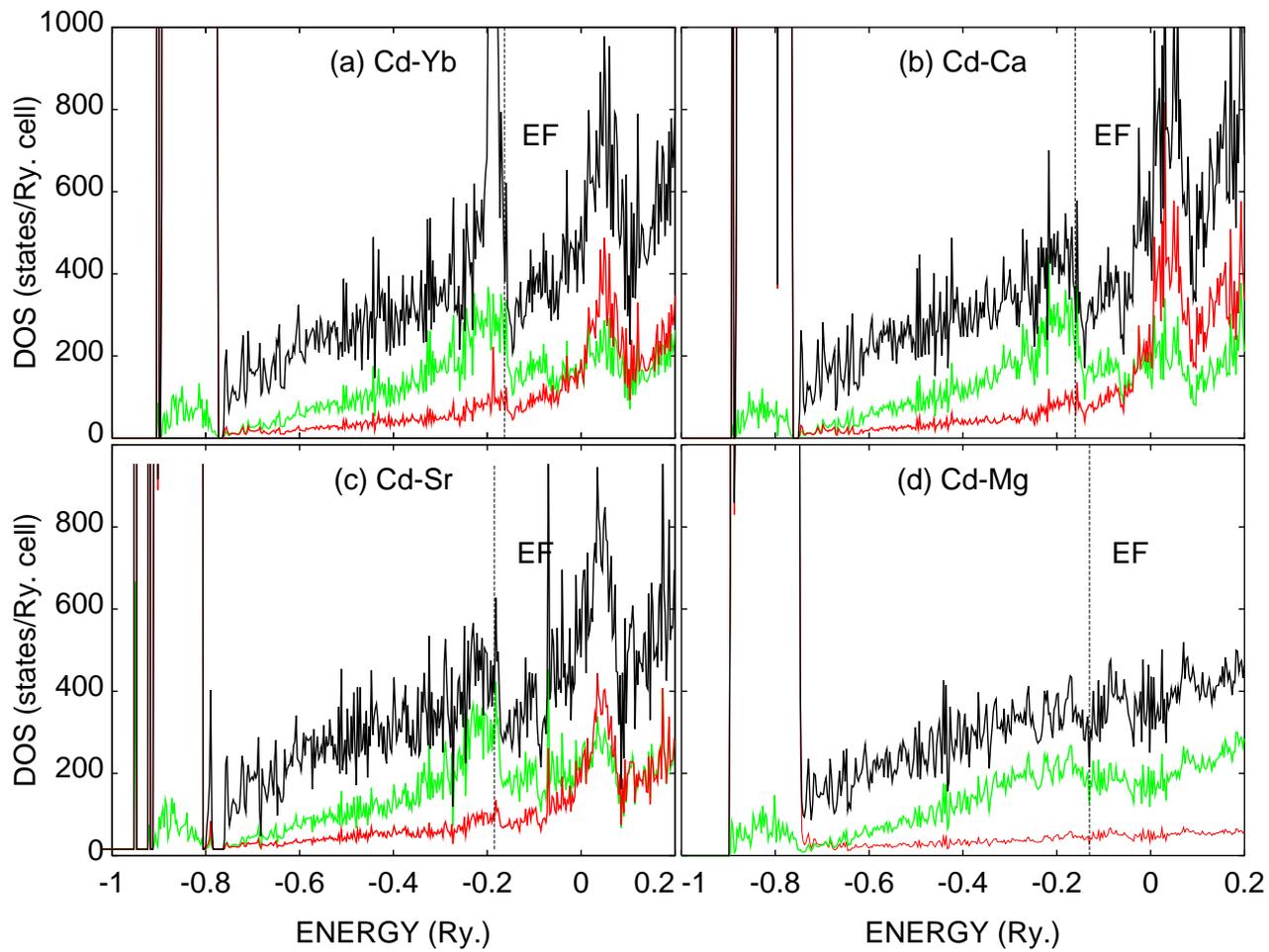
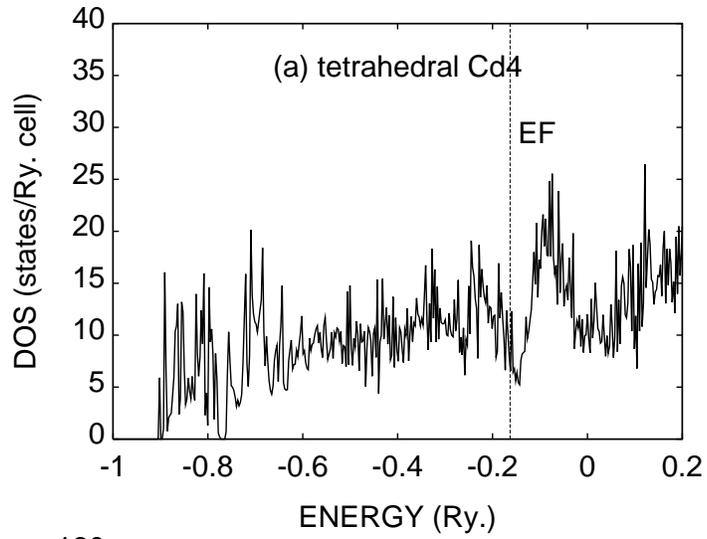
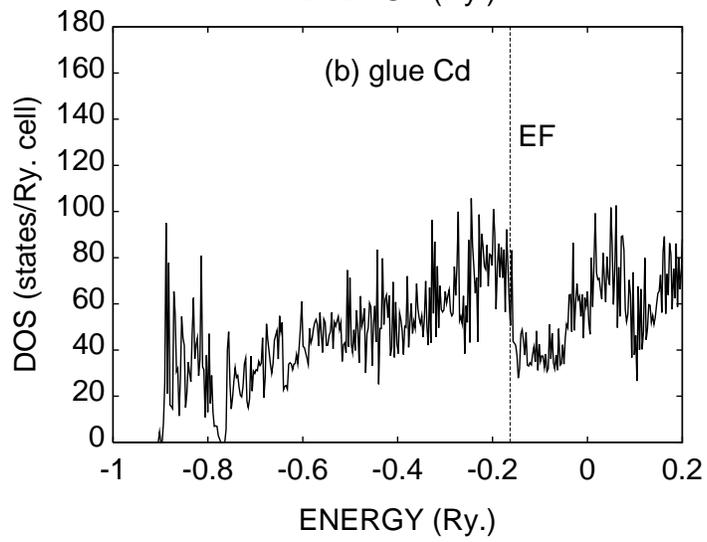
\end{document}